\documentclass[preprint2]{aastex}
\begin{document}

\title{The Megaparsec-Scale X-ray Jet of the BL Lac Object OJ287}

\author{Alan P. Marscher\altaffilmark{1} and Svetlana G. Jorstad\altaffilmark{1,2}}
\altaffiltext{1}{Institute for Astrophysical Research, Boston University,
    725 Commonwealth Ave., Boston MA 02215}
\altaffiltext{2}{Astronomical Institute, St. Petersburg State University, Universitetskij Pr. 28, 
Petrodvorets, 198504 St. Petersburg, Russia}

\begin{abstract}
We present an X-ray image of the BL~Lacertae object OJ287 revealing a long jet, curved by
$55^\circ$ and extending $20''$, or 90~kpc from the nucleus. This de-projects to $>1$~Mpc based on
the viewing angle on parsec scales. Radio emission follows the general X-ray morphology but
extends even farther from the nucleus. The upper limit to the isotropic radio luminosity,
$\sim2\times10^{24}$~W~Hz$^{-1}$, places the source in the Fanaroff-Riley~1 (FR~1) class,
as expected for BL~Lac objects. The spectral energy distribution indicates that the extended
X-ray emission is from inverse Compton scattering of cosmic microwave background photons.
In this case, the derived magnetic field is $B\approx5~\mu$G, the minimum electron energy is
7-40$m_{\rm e}c^2$, and the Doppler factor is $\delta\approx8$ in a knot $8''$ from
the nucleus. The minimum total kinetic power of the jet is 1-2$\times10^{45}$~erg~s$^{-1}$.
Upstream of the bend, the width of the
X-ray emission in the jet is about half the projected distance from the nucleus. This implies that
the highly relativistic bulk motion is {\it not} limited to an extremely thin spine, as has been
proposed previously for FR~1 sources. The bending of the jet, the deceleration of the flow
from parsec to kiloparsec scales, and the knotty structure
can all be caused by standing shocks inclined by $\sim 7^\circ$ to the jet axis. Moving shocks
resulting from major changes in the flow properties can also reproduce the knotty structure,
but such a model does not explain as many of the observational details.
\end{abstract}

\section{Introduction}

According to the unified scheme of extragalactic radio sources, BL Lacertae (BL Lac) objects
represent Fanaroff-Riley~1 \citep[FR~1;][]{FR74} radio galaxies in which one of the jets
points within $\sim 15^\circ$ of the line of sight \citep{AU85,PU90}.
Based on a careful analysis of the radio jets of 38 radio
sources, \citet{L99} inferred that the bulk flow of a typical FR~1 jet decelerates from a
relativistic speed to less than $\sim 0.1c$ within a distance of two to tens of kpc from the
nucleus. The inferred deceleration agrees with the expectations of a model in
which the jet entrains material from its surroundings \citep{Bick94,L96}. The distance
from the nucleus over which the deceleration occurs is then expected to increase with the kinetic
power of the jet. It should therefore be greater at higher values of both the initial bulk
Lorentz factor and the luminosity of extended emission.
Laing et al. (1999) also found that the data are consistent with a model
in which the velocity of the flow decreases
from the axis to the edge of the jet, either gradually or in a more distinct manner, as
a spine plus sheath. \citet{Ch00} came to the same conclusion based
on a comparison of BL Lac and FR~I core spectral energy distributions. Some jet
launching models include an ultra-fast spine produced by the ergosphere of a rotating
black hole, with the slower sheath emanating from the accretion disk \citep[e.g.,][]{P96,McK06}.
Alternatively, \citet{Meier03} proposed that the spine is generated by the central
engine, while the sheath arises from velocity shear downstream.

The proposed deceleration of jets could
reconcile gross discrepancies between inferred bulk Lorentz factors $\Gamma$ of the
BL Lac and FR~I populations. \citet{U91} concluded that the properties of BL Lac objects
require the values of $\Gamma$ of the FR~I parent population to range up
to $\sim 30$. There is independent support for this conclusion: measurements
of proper motions of $\gamma$-ray bright BL Lacs by \citet{J01}
indicate apparent speeds (and therefore minimum values of $\Gamma$) as high as $\sim 40c$.

The extended X-ray emission from low-power FR~1 radio galaxy jets is
consistent with synchrotron radiation within the inner 10 kpc by electrons with
energies in the 10-100 TeV range \citep{W01,KS05}. Those that are more luminous,
highly relativistic, and pointing close to the line of sight --- characteristics of radio-bright
BL Lac objects --- can also produce observable X-rays from inverse Compton scattering off the
microwave background \citep[IC/CMB;][]{Tav00,Cel01,Sam01}
on scales of tens or hundreds of kiloparsecs. The total energy requirements are higher in
the IC/CMB case owing to the radiative inefficiency of low-energy electrons \citep{DA04}.
This has led to a controversy over the process by which X-ray emission is produced in
the jets of quasars and BL Lac objects with highly relativistic jets.

Until now, observations that provide information on deceleration, spine-sheath structure,
and X-ray emission have been limited to FR~1 radio galaxies that are not BL Lac objects
\citep[e.g.,][]{W01}, three BL Lac objects with mixed FR~1 and FR~2 morphologies \citep{Sam07,Sam08},
and blazar-class quasars with powerful, highly relativistic jets \citep[e.g.,][]{JM04,HK06}.
The extended X-ray emission of low-power, high-$\Gamma$ BL Lac objects has been less well
studied. Such jets should radiate at X-ray energies via the IC/CMB and/or synchrotron
process, with the emission gradually becoming weaker with distance from the nucleus as
the jet decelerates. Because of the low power of the jet as determined by the luminosity of
the extended emission, the IC/CMB process can be eliminated as a major contributor to the
X-ray emission if the total energy requirement is too high, as is the case for some
quasars \citep{DA04}. Furthermore, if the jet possesses a spine-sheath structure,
the IC/CMB emission should arise only in the narrow, highly relativistic spine.

In order to explore the low-power, high-$\Gamma$ case, we have carried out multi-waveband
imaging of the jet of the BL Lac object OJ287 with the {\it Chandra} X-ray Observatory (CXO),
{\it Hubble} Space Telescope (HST), {\it Spitzer} Space Telescope (SST), and Very
Large Array (VLA). We report here the detection of X-ray emission along a length of the
jet extending $\sim 90$ kpc from nucleus in projection, and more than 1 Mpc after
de-projection. The X-ray emission, which has a somewhat knotty structure, is resolved in the
transverse direction and fades gradually with distance from the nucleus, as does the radio
emission. The radio and X-ray fluxes, along with upper limits at infrared and optical
wavelengths, provide a spectral energy distribution that strongly constrains the emission
model.

We discuss the observations in {\S}2 and analyze the results in {\S}3. {\S}4 contains
our interpretation of the observations and discusses the implications for extended jets, while
{\S}5 presents our main conclusions. In this study, we adopt the current standard
flat-spacetime cosmology, with Hubble constant $H_0$=71 km s$^{-1}$ Mpc$^{-1}$,
$\Omega_M=0.27$, and $\Omega_\Lambda=0.73$. At a redshift $z=0.306$, OJ287 has a
luminosity distance $d_\ell$ = 1.58 Gpc, and $1''$ corresponds to a projected distance
of 4.48 kpc in the rest frame of the host galaxy of OJ287.

\section{Observations and Data Analysis}

Table \ref{tab1} lists the main information about the CXO, HST, SST, and VLA observations of
OJ287. The VLA data include both a new observation at a wavelength of 2 cm and archival data
at 6 and 21 cm.

\subsection{X-Ray Observations}

\begin{figure}
\epsscale{1.0}
\plotone{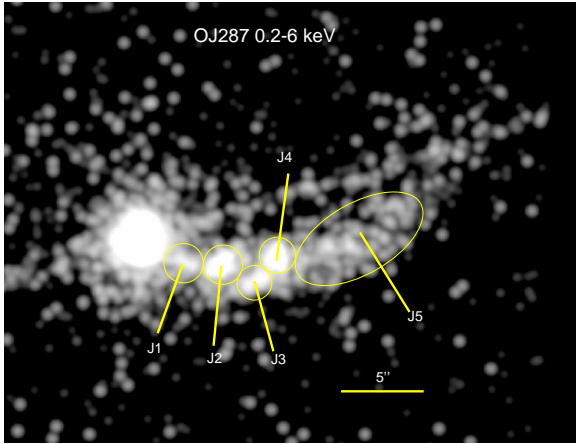}
\caption{CXO ACIS-S3 X-ray image of OJ287 at 0.2-6~keV convolved with a Gaussian kernel of
FWHM$=0''\kern -0.35em .5$. The X-ray knots are labeled. }
\label{fig1}
\end{figure}

We observed OJ287 using the back-illuminated S3 chip of the Advanced Camera for Imaging and
Spectroscopy (ACIS) on CXO and a 1/8 subarray to reduce the frame 
time to 0.4~s. The data analysis, which utilized version 3.3.0 of the CIAO software and
version 3.2.4 of {\it CALDB}, followed the procedures detailed in \citet{JM04}.
We generated a new level 2 event file with a pixel size of $0''\kern -0.35em .1$ 
and convolved the image with a $0''\kern -0.35em .5$ FWHM circular Gaussian kernel corresponding
to the angular resolution of the ACIS.
The resulting image (Fig. \ref{fig1}) shows a prominent jet that extends out to $\sim 20''$
from the nucleus. The direction of the jet changes by $55^\circ$ from the straight inner
section within $8''$ of the core to the end of the X-ray emission. The jet has a knotty
structure, which we separate into five regions, $J1-J5$. All knots are detected at a level
of 10$\sigma$ or higher relative to the background.

We fit the spectral data for the jet in the photon energy range 0.2-6~keV 
by a power-law model with fixed Galactic absorption $N_H=3.02\times 10^{20}$ cm$^{-2}$
\citep{DL90}. (Allowing the column density to vary yields a value that is not significantly different
from this.) The calculated parameters of the X-ray jet are given in Table \ref{tab2}. 

\subsection{Optical Observations}

\begin{figure}
\epsscale{1.0}
\plotone{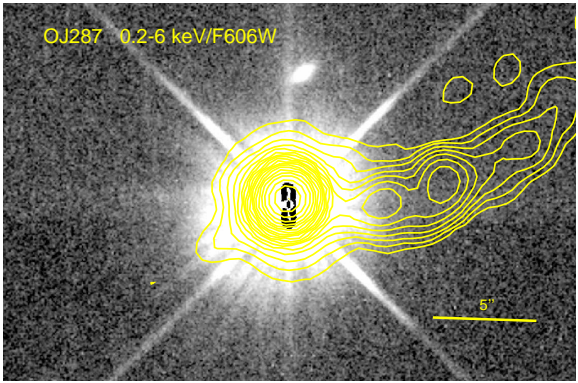}
\caption{HST optical image of OJ287 (gray scale) after PSF subtraction, with X-ray (yellow) contours superposed. Contours are in factors of 2 starting at 2.075 erg s$^{-1}$ cm$^{-2}$ beam$^{-1}$,
where the convolving beam is the same as for Figure \ref{fig1}. The image is rotated by
$\sim 17^\circ$ clockwise from that of the images in the other figures.}
\label{fig2}
\end{figure}

Over two orbits of HST, we obtained 12 exposures of 230~sec each of OJ287 with the WFPC2 PC-chip
using the broad filter F606W in a 4-point dithering mode. The images were combined with task
crrej within the STSDAS software package. We generated the point-spread function (PSF)
with the TinyTim software and subtracted it from the combined image.
The deconvolved image is displayed in Figure~\ref{fig2}. We did not detect significant flux over
the sky level along the position of the X-ray and radio jet. \citet{Y97} also did not detect
the jet at 814 nm with the WFPC2, and found that emission from nebulosity associated with
the host galaxy is insignificant beyond $4''$ from the nucleus.

\subsection{Infrared Observations}

\begin{figure}
\epsscale{1.0}
\plotone{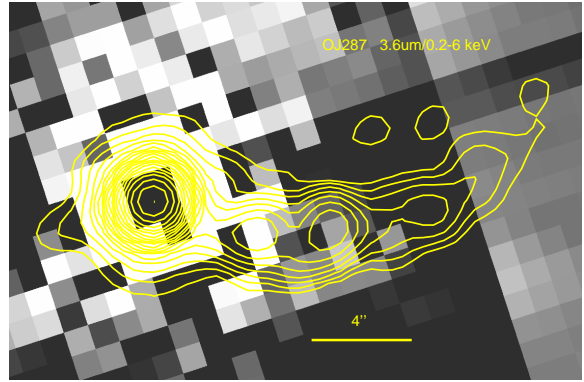}
\caption{SST image at $3.6\mu$m of OJ287 (gray scale) after PSF subtraction, with X-ray (yellow)
contours  superposed. Contour levels are as in Figure \ref{fig2}. }
\label{fig3}
\end{figure}

We observed OJ287 with IRAC on SST at 3.6 and 5.8~$~\mu$m with a single frame exposure of 12~s
in a dithering mode (36 Positions Reuleaux). We used the Level 2 Post-BCD Pipeline images,
which we checked for artifacts. We created the PSF at each infrared (IR) wavelength using
image stars, then
subtracted it from the image  with the task apex\_qa\_1frame.pl of version 18.3.1 of the
MOPEX/APEX software package. Figure~\ref{fig3} displays the deconvolved image at 3.6~$\mu$m,
which reveals some enhanced brightness in the jet direction, especially in the regions of knots
$J2$ and $J4$. However, the fluxes are at 1-2$\sigma$ detection level owing to possible
confusion with unrelated IR sources, hence we cannot ascertain that the emission is
physically associated with the jet. We therefore use the measured fluxes (plus the 2-$\sigma$
uncertainties) to estimate the IR upper limits for the jet regions.

\subsection{Radio Observations}

\begin{figure}
\epsscale{0.8}
\plotone{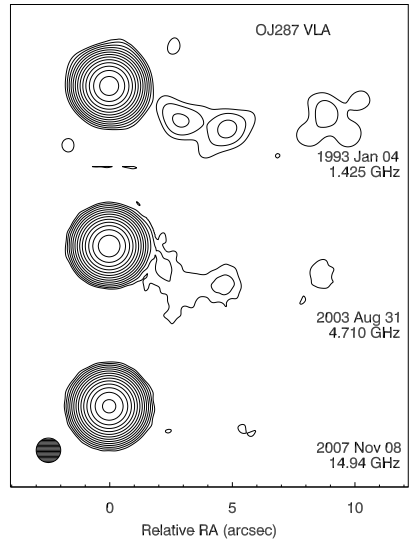}
\caption{VLA images of OJ287. The global total intensity peak is 2.87~Jy/beam at 14.94~GHz.
The images are convolved with a circular Gaussian beam of dimensions 1$\times$1 arcsec$^2$.
Contours are in factors of 2 starting at 0.01\% of the global peak intensity of
2.87 Jy beam$^{-1}$ (reached at 14.94 GHz).
\label{fig4}}
\end{figure}

We processed the VLA data with the Astronomical Image Processing System (AIPS) software 
provided by the National Radio Astronomy Observatory (NRAO), following the standard procedures
outlined in the {\it AIPS Cookbook} available at www.nrao.edu. (The observations at 15 GHz
included only 15 of the usual 27 antennas, since they occurred during the
changeover to the new EVLA.) Each observation employed two 50 MHz bands, centered on
1.3851, 1.4649 GHz (21 cm wavelength), 4.5351, 4.8851 GHz (6 cm), and 14.915, 14.965 GHz
(2 cm), with mean frequencies listed in Table \ref{tab1}. We performed the imaging
with the Difmap software \citep{Difmap}, again following standard procedures. Figure~\ref{fig4}
presents all three radio images, while Table \ref{tab2} gives the parameters of the radio emission.
The image at 1.425 GHz is very similar to the full-resolution version presented by \citet{PS94},
which was based on the same data.

\section{Observational Results}

\begin{figure}
\epsscale{1.0}
\plotone{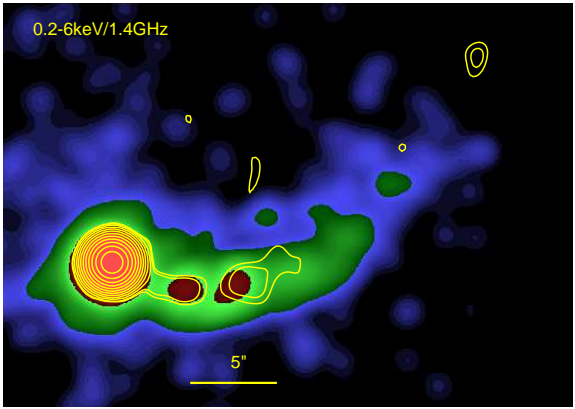}
\caption{Superposed images of OJ287, smoothed to enhance the large-scale
structure of the jet: X-ray (0.2-6~keV, false-color, smoothed with
a FWHM$=1''\kern -0.35em .5$
circular Gaussian function) and radio (21 cm, contours, with circular Gaussian restoring beam of
FWHM$=1''\kern -0.35em .5$ approximately corresponding to natural
weighting of the VLA fringe-visibility data).
\label{fig5}}
\end{figure}

Figure \ref{fig5} displays the entire jet at both radio and X-ray frequencies. The
radio and X-ray emission is generally coincident within $12''$ of the nucleus, beyond which
the radio emission falls below the detection limit. An exception is knot J3, situated
along the southern rim of the jet with centroid $7''\kern -0.35em .5$ from the nucleus,
where we detect no radio emission above the noise level. Because of the low radio flux, we
cannot reliably determine the flux density separately at 4.7 or 14.9 GHz
for each knot denoted in Figure \ref{fig1}. We therefore measure these quantities only
for the entire extended jet from knot J1 to J5, with the values given in Table \ref{tab2}.

\subsection{Comparison of X-ray and Radio Emission}

\begin{figure}
\epsscale{1.0}
\plotone{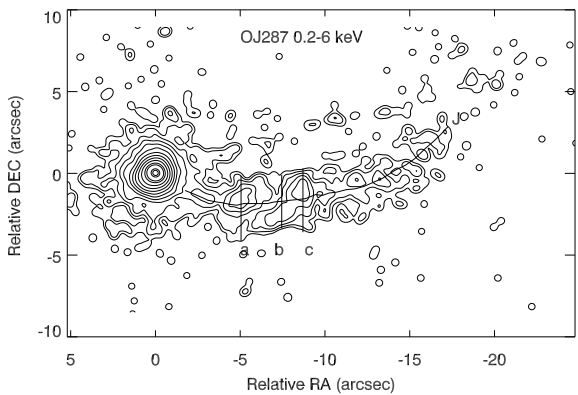}
\caption{Contour map of X-ray intensity, with curves and lines indicating the paths corresponding
to the intensity profiles of Figure \ref{fig7}. The X-ray image is smoothed with a
circular Gaussian restoring beam with FWHM size $0''\kern -0.35em .5$, while the contour levels
are as in Figure \ref{fig2}. }
\label{fig6}
\end{figure}

\begin{figure}
\epsscale{1.0}
\plottwo{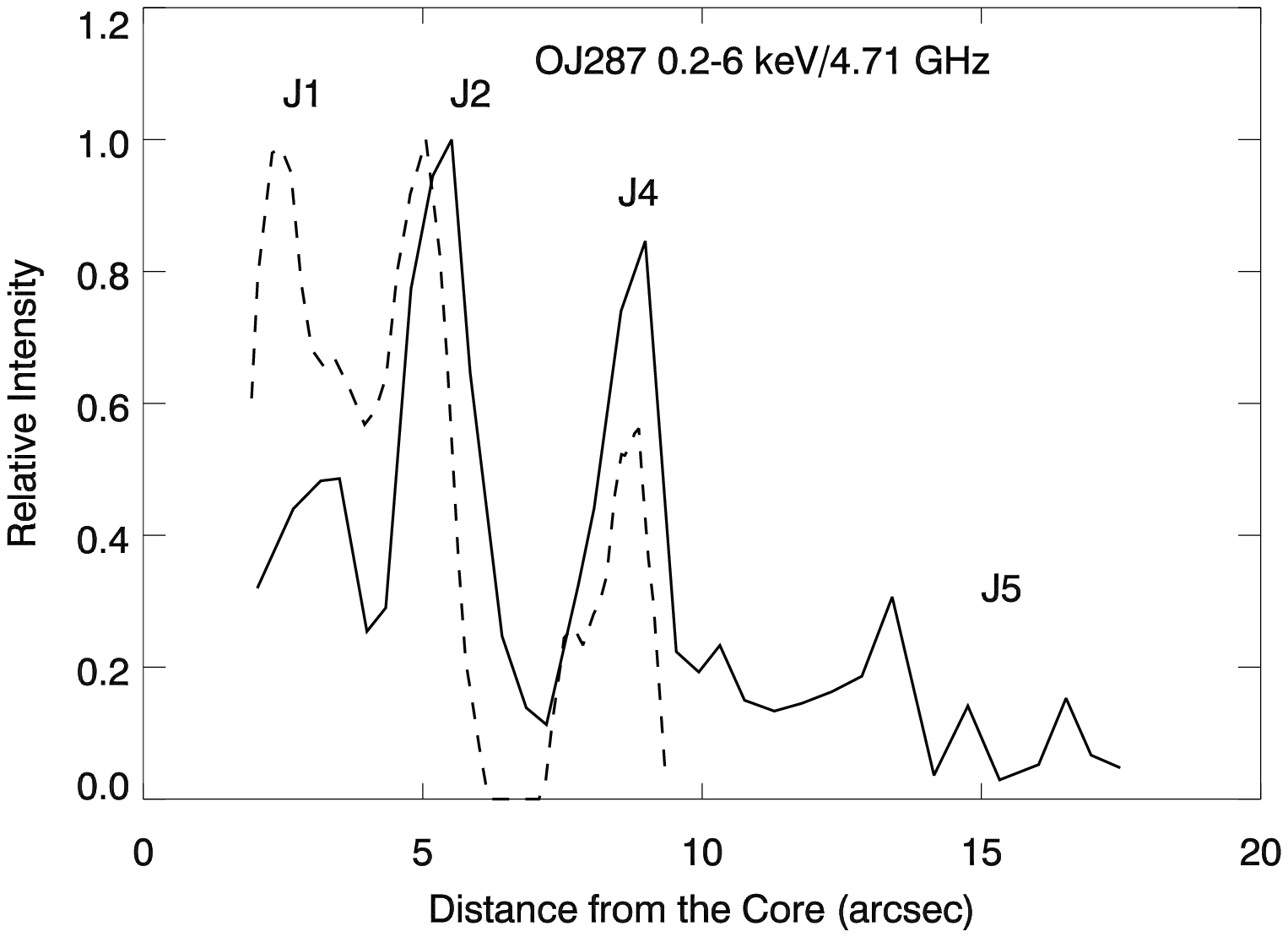}{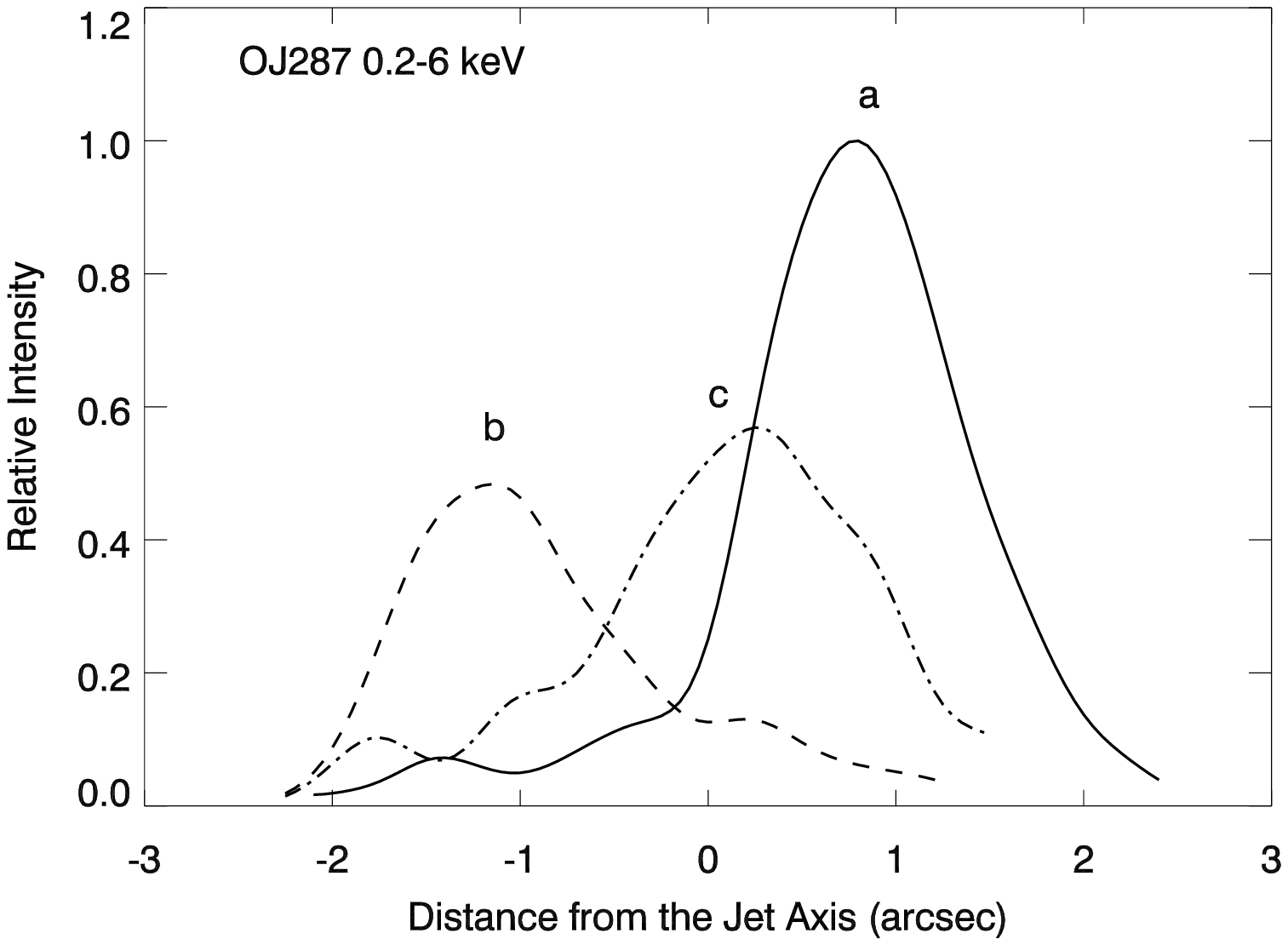}
\caption{Intensity profiles along the paths marked in Figure \ref{fig6}. {\it Left:}
X-ray intensity (solid curves) and radio intensity at 4.7 GHz (dashed curves)
along the jet axis. {\it Right:} X-ray intensity transverse to the axis, proceeding from
south to north. }
\label{fig7}
\end{figure}

Figure \ref{fig6} marks the jet axis (which traces the mid-point of the detected X-ray
emission transverse to the axis) and three crosscuts that pass through knots $J2$, $J3$,
and $J4$. In Figure \ref{fig7} we plot both
the X-ray and radio intensity profiles along the jet axis, and only the transverse X-ray
profile along the crosscuts. The
path along the axis misses knot J3, which lies along the southern boundary of the jet,
but crosscut {\it b} includes the knot. Knots J1, J2, and J4 are clearly seen in the
axial profile, while crosscut {\it a} passes through knot J2 and {\it c} includes J4.
The maximum X-ray intensity lies farther from the nucleus than does the radio peak
in both J1 and J2 by
$0''\kern -0.35em .9$ and $0''\kern -0.35em .5$, respectively, while the peaks
at the two wavebands coincide in knot J4. In each case, however, the upstream boundaries
of the radio and X-ray emission coincide.

The displacement of the intensity peaks of knots J2-J4 from the axis (defined as the
centroid of the detectable X-ray emission across the jet) seen in Figures \ref{fig1}
and \ref{fig7} is striking. On the other hand, the intensity along the axis is too
strong to consider the jet to be edge-brightened.

The spectral index of the radio emission from the jet, integrated over the region containing
detectable X-ray flux, is $\alpha$(radio)$ = 0.8\pm 0.1$. This differs somewhat from the X-ray
value, $\alpha$(X-ray)$ = 0.61\pm 0.06$. The discrepancy could be due to a gradient in X-ray
spectral index, which increases with distance from the nucleus from knots J2 to J4. We note
that the overall radio spectral index falls within the 1-$\sigma$ range of the X-ray spectral
index for all knots except J2. Given this, and because we can only determine the overall
radio spectral index, we will ignore the small apparent radio/X-ray discrepancy in spectral slope
and adopt a value $\alpha=0.7$ in our analysis.

The radio emission on milliarcsecond scales consists of a bright ``core'' with multiple knots ---
some with apparent superluminal motion and others that are subluminal or even stationary ---
lined up along position angles that vary with time. These cover a range over more than $90^\circ$
\citep{TK04,Agudo10}, which is amplified by projection effects owing to the small angle of
the compact jet to the line of sight, $3.\kern -0.35em ^\circ 2\pm 0.\kern -0.35em ^\circ 9$
\cite{J05}. The mean over a 7.5-year period was $-107^\circ$ (measured from
north, with $-90^\circ$ defined as due west) \citep{TK04}, while the mean position angle of
the X-ray jet within $8''$ of the nucleus is $-109^\circ$. This implies that, independent
of the direction of motion on parsec scales, the flow bends to collimate along a
well-defined direction on kiloparsec scales, as has been observed directly in 3C~279 \citep{H03}.

\subsection{Upper Limit to Unbeamed Radio Luminosity}

The parsec-scale superluminal motion, with apparent velocities as high as $18c$ \citep{J05},
indicates that bulk motions in the jet are highly relativistic inside the nucleus.
\citet{J05} combined the apparent speeds with the rate of decline of flux of moving knots
to derive a Doppler factor $\delta = 18.9\pm6.4$, a bulk Lorentz factor
$\Gamma = 16.5\pm4.0$, and an angle between the axis and line of sight of
$3^\circ \kern -0.35em .2\pm0^\circ \kern -0.35em .9$ for the compact jet. The one-sidedness
of the extended X-ray and radio emission implies that relativistic bulk motion continues at
least out to an angular distance of $20''$ from the nucleus.

We cannot determine whether {\it any} of the radio emission that we detect is unbeamed. However,
given the limit to the sensitivity of our radio images, it is possible that the two
outermost features seen in Figure \ref{fig5} are located in a very faint lobe or other region
where there is only non-relativistic bulk flow. We measure a flux density at 1.425 GHz from
this region to be $4.6\pm0.9$ mJy, which gives a 2-$\sigma$ upper limit to the
unbeamed extended radio power at 1.4 GHz of $< 2\times10^{24}$ W Hz$^{-1}$. We therefore
confirm that this BL Lac object is within the luminosity range expected for FR~1 radio
sources \citep{FR74}.

\subsection{Spectral Energy Distribution}

\begin{figure}
\epsscale{1.0}
\plotone{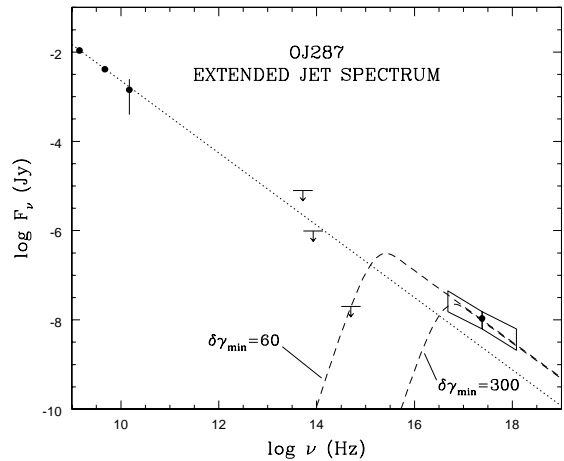}
\caption{Continuum spectrum of OJ287. The X-ray data are displayed by the flux density and
uncertainty at $2.4\times 10^{17}$ Hz (1 keV), with solid lines corresponding to
$F_\nu \pm$1-$\sigma$ at other frequencies. The dotted line gives the extrapolation of
the radio spectrum. The dashed lines correspond to the spectrum expected for inverse Compton
scattering of CMB photons for the two extreme values of the Doppler factor times the
minimum electron energy that fit the X-ray and optical data.}
\label{fig8}
\end{figure}

Figure \ref{fig8} presents the multifrequency continuum spectrum of the
extended jet. The measurements include the entire length of the detected extended
X-ray jet, i.e., from $3''$ to $20''$ from the nucleus in order to provide a signal-to-noise
ratio at 4.7 and 14.9 GHz sufficiently high for analysis and to allow comparison with
the low-resolution IR images. The figure reveals that the X-ray emission lies slightly above an
extrapolation of the radio spectrum, while the 3.6$~\mu$m IR and 600 nm optical upper
limits fall well below the extrapolation. In fact, the 600 nm upper limit is comparable
to the X-ray flux measurement. This is contrary to models in which the X-rays arise
from synchrotron radiation from electrons with energies in the TeV range. In such
models \citep{DA02,DA04,J06,U06}, the value of $\nu F_{\nu}$(X-ray) can exceed
$\nu F_{\nu}$(opt), but the flux densities cannot unless the electron energy spectrum is
extraordinarily shallow, $N(\gamma) \propto \gamma^{-s}$ with $s \approx 1$, where
$\gamma$ is the electron energy in units of $m_{\rm e}c^2$. In this case, essentially all of the
electron energy would be confined to TeV energies. Furthermore, the spectral index in
the X-ray region should then be $\alpha{\rm (X-ray)} \sim 0$, while our measured value is
significantly higher.

Having eliminated the synchrotron model, we construct the SED for the alternative model,
IC/CMB, by integrating over both the electron energy distribution [assumed to be of the form
$N(\gamma) = N_0\gamma^{-(2\alpha+1)}$] and the spectrum of the
CMB to calculate the X-ray flux density as a function of photon energy. In order to avoid
violating the upper limit at 600 nm, we cut off the electron energy distribution below
an energy $\gamma_{\rm min}$. We present two SEDs for this model in Figure \ref{fig8},
representing the lowest and highest allowed values of the product $\delta \gamma_{\rm min}$
that provide good fits to the observed SED.

\section{Discussion}

\subsection{Physical Parameters of the Extended Jet}

We use the formulas derived in the Appendix of \cite{JM04} to determine the physical parameters
of the extended jet of OJ287 under the assumptions that the X-ray emission is from the
IC/CMB process and that the magnetic and particle energy densities are equal, with
random kinetic energies of the protons and electrons each contributing 25\% of the total
energy density. We do this for knot J4, $8''$ from the nucleus,
which is well resolved from the other features
in both the X-ray and radio images. The knot is nearly circular (see Fig.\ \ref{fig1}),
with a mean extent of $a = 1''\kern -0.35em .4$. We assume that the radio and X-ray spectral
indices are the same, and carry out the calculations for the value that is most consistent
with the data at both wavebands, $\alpha = 0.7$.

The above values of $a$ and $\alpha$, plus the flux densities listed in Table
\ref{tab2} yield a Doppler factor $\delta = 8.7$ and magnetic field $B = 5.4 ~\mu$G for the
case $\gamma_{\rm min} = 7$ that represents the leftmost curve in Figure \ref{fig8}, while
for the rightmost curve $\delta = 7.9$, $B = 4.9 ~\mu$G, and $\gamma_{\rm min} = 40$.
In order to derive the minimum total jet power, we assume equipartition between the magnetic and
electron energy densities.  We then multiply the magnetic energy density $B^2/8\pi$ by
$4\pi$ times the square of the luminosity distance, the cross-sectional area $\pi (a/2)^2$ and speed
($\approx c$) of the jet, and the square of the bulk Lorentz factor (with the
approximation that $\Gamma \approx \delta$) to transform to the host galaxy rest frame.
We double the result, since the magnetic field supplies only half of the total energy under
the equipartition assumption, to arrive at a total power of
1-2$\times 10^{45}$ ergs s$^{-1}$. This is similar to that derived from a detailed analysis
of very long baseline interferometric images for the parsec-scale jet of the FR~1 radio
galaxy 3C~120 by \citet{Mars07}. Both analyses, however, ignore the rest mass of the
protons, which multiplies the derived kinetic power by $\sim 1800 f \gamma_{\rm min}^{-1}$,
where $f$ is the fraction of the positively charged species that are protons rather than
positrons. The total kinetic power is therefore $\sim 1\times 10^{45}(1+50f)$ ergs s$^{-1}$
for the value $\gamma_{\rm min} = 40$ that minimizes the power requirement. This corresponds to
$\sim 3 M_{\rm BH,8}^{-1}$ times the Eddington luminosity of a black hole with mass
$1\times 10^8 M_{\rm BH,8}$~$M_\odot$. If the mass of the black hole that powers OJ287
is as high as $1.8\times 10^{10}$~$M_\odot$, as in the binary model of \citet{Val08},
then the total kinetic power is much less than the Eddington value. The jet power
can be sub-Eddington even if the black-hole mass is $\sim 10^9$~$M_\odot$.
We therefore find that the EC/CMB model for the X-ray emission produces reasonable physical
parameters, especially if the value of $\gamma_{\rm min}$ near the maximum of $\sim 40$
allowed by the SED. 

\subsection{Bending, Length, and Deceleration of the Jet}

We note that the derived value of the Doppler factor in knot {it J4} $8''$ from the nucleus is
a factor of 2-3 lower than that found in the parsec-scale jet \citep{J05}. This implies
that the jet flow either decelerates on scales of hundreds of kiloparsecs or varies over
time-scales of thousands of years.
The main bending of the jet appears to occur beyond $8''$ from the nucleus. If the jet
remains as straight in three dimensions as it appears projected on the sky, as is the
most likely case, then we can adopt the same value for the angle of the axis to the line
of sight as that found on parsec scales,
$\theta = 3.\kern -0.35em ^\circ 2\pm 0.\kern -0.35em ^\circ 9$
\citep{J05}. The deprojected length of the straight portion of the jet is then
$640^{+940}_{-140}$ kpc. In order for a jet at an initially small viewing angle to appear to
bend by more than that angle, it must curve in three dimensions by at least the same angle
as $\theta$. Such bending should require a
shock at the boundary of the jet opposite the side toward which it curves \citep{CF48}.
Knot J3 could represent such a shock. If the total curvature of the jet in three
dimensions is near the minimum required to produce the apparent bending by $55^\circ$,
$\Delta\theta \gtrsim 3^\circ\kern -.35em .2$,
then the final viewing angle increases to $\sim 7^\circ$. With these values, we estimate the
total deprojected length of the X-ray jet to be $\gtrsim 1$ Mpc, with the radio jet extending
even farther.

The Doppler factor that we have obtained from our analysis is still quite high, indicating
that the jet flow remains highly relativistic out to hundreds of kiloparsecs from the nucleus.
This is contrary to the deceleration to sub-relativistic speeds and disruption on scales of
tens of kiloparsecs that has been inferred to be typical of
FR~1 radio galaxies \citep{L96,L99}. However, OJ287 cannot be typical, given the
highly relativistic nature of the jet on parsec scales, with $\Gamma = 16.5\pm4.0$,
compared with a more common value closer to $\Gamma \sim 1$
\citep[see][]{LM97}. The extension of the jet beyond 1 Mpc therefore agrees with the
theoretical expectation (see {\S}1) that more highly relativistic (and presumably more highly
supersonic) flows can penetrate the interstellar and intergalactic medium out to greater
distances before being slowed down and disrupted.

\subsection{Knotty Structure of the Jet and Shock Models}

The bulk of the extended X-ray emission seen in Figure \ref{fig1} emanates from several
bright knots. Figure \ref{fig7} shows that the knot to inter-knot flux ratio is as high
as $\sim 10$:1 in X-rays and even greater at 4.7 GHz. A successful emission model must
be consistent with this morphology. In fact, such knotty structure has been
used as an argument against the IC/CMB
model on the grounds that the energies of the electrons that scatter the CMB photons to
the X-ray range are low, $\gamma \sim 40$-130 in the case of OJ287. Because of this, the
electrons have lifetimes to radiative energy losses that are too long to confine the
emission to knots of length 5-10 kpc \citep{Sta04}. This is true even for electrons
radiating at centimeter wavelengths, whose Lorentz factors are quite high,
$\gamma \sim 10^4$ for magnetic fields in the several micro-Gauss range. The radiative-loss
lifetimes (mainly from IC/CMB radiation) of these electrons are $\gtrsim 10^6$ yr.
While gradients in magnetic field can cause synchrotron intensity contrasts, the knot to
inter-knot flux ratio is too high for this to be the sole cause. The knotty structure
must therefore involve another process. We now consider hydrodynamical effects that
can potentially produce emission features with sizes $\lesssim10$ kpc.

\subsubsection{Standing Shock Model}

Hydrodynamical effects that occur in the flows of jets \citep{CF48} can cause spatial
variations in both density and bulk velocity.
Interactions with the external medium form shocks oblique to the jet axis that decelerate,
compress, and bend the jet flow \citep{Gom97}, as well as energize relativistic particles.
In between the shocks are rarefactions,
which increase the Lorentz factor $\Gamma$ at the expense of lower energy
density $u_{\rm e} \propto n_{\rm e}\gamma_{\rm min}$ of the electrons, where $n_{\rm e}$ is
the number density of electrons. On the other hand the energy density of relativistic
electrons is inversely related to $\Gamma$; e.g., in the case of adiabatic compression and
rarefactions and an ultra-relativistic equation of state,
$\Gamma \propto u_{\rm e}^{-1/4} \propto [N_0\gamma_{\rm min}^{(1-2\alpha)}]^{-1/4}$.
Because the IC/CMB intensity depends on these factors as
$I_{\rm IC/CMB} \propto N_0\delta^2\Gamma^2 \sim N_0\Gamma^4$
for a small viewing angle, the $\Gamma \propto N_0^{-1/4}$ relationship would limit the
knot to inter-knot intensity ratio to a value near unity if compression were the only
physical effect. We therefore need to include non-adiabatic effects of shocks to explain the
presence of distinct knots of IC/CMB emission.

We use the formulas given in \citet{LB85} and \citet{CC90} to calculate $\Gamma$ and the angle of
deflection $\xi$ of the flow velocity behind the shock, as well as the ratio of post-shock
to pre-shock densities. These depend on the upstream value of $\Gamma$
and the angle of inclination $\eta$ of the shock front relative to the upstream flow
direction. We consider the most extreme case, in which $\Gamma_{\rm up} = 16.5$, as derived
on parsec scales, is adopted for the upstream flow and $\Gamma_{\rm down} \approx 8$ downstream of
the shock front, as in knot {\it J4}. This occurs for $\eta =  7^\circ$ and causes a deflection
$\xi = 4^\circ$ that is adequate for a large bend to appear in projection,
as observed. We note that, since it is the velocity component that is transverse to the
shock front that decreases, the shock must be oriented in the direction of the bend,
which can be seen to be the case for knot {\it J3} in Figure \ref{fig1}. We calculate the upstream
to downstream density contrast to be a factor of 5 in this model. If we include compressional
heating by the shock, so that $\gamma_{\rm min} \propto n_{\rm e}^{1/3}$, we find that
$N_0 \propto n_{\rm e} \gamma_{\rm min}^{2\alpha} \propto n_{\rm e}^{(3+2\alpha)/3}$ increases by
a factor of 11 from the compression for our adopted value $\alpha=0.7$. However,
the decrease in $\Gamma$ across the shock offsets this when calculating the expected IC/CMB
intensity, since $I_{\rm IC/CMB} \propto N_0\Gamma^4$. In order to
explain the intensity contrast, the shock would need to heat the electrons more strongly
by another process, probably diffusive particle acceleration \citep[e.g.,][]{EBJ96,MQ03}.
In order to bring the knot to inter-knot intensity ratio to the observed value of $\sim 10$ by
increasing $N_0$, the minimum (and therefore the mean) relativistic electron energy would
need to increase by a factor of $\sim 16$, i.e., 8 times more than by compression alone.
The required increase in $\gamma_{\rm min}$ could be reduced by a factor of 1.6
if $\Gamma_{\rm up}$ were 10 (roughly the minimum Lorentz factor that is compatible with
a shock having $\Gamma_{\rm down}=8$) instead of the parsec-scale value of 16.5.

\begin{figure}
\epsscale{1.0}
\plotone{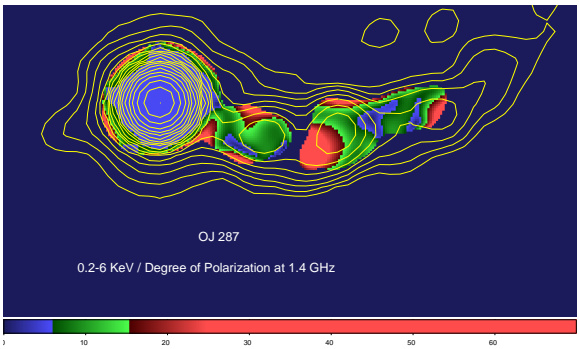}
\caption{Map of degree of linear polarization of OJ287 at 1.425 GHz. The superposed X-ray contours
are the same as in Figure \ref{fig2}. The scale on the bottom is in percent.}
\label{fig9}
\end{figure}

Since bending and deceleration of the jet near the X-ray knots can be caused
by interactions with the external medium that set up standing oblique shocks that deflect
the flow, a shock model can potentially explain both features of the jet in OJ287.
This conclusion is supported by the crude 1.425 GHz polarization map displayed
in Figure \ref{fig9}. (Unfortunately, the noise level of the polarized intensity is too high
even to estimate the position angle of the polarization after statistical bias is
taken into account.) The highest polarization is between knot J3 and the southeastern
side of J4, where the jet first bends noticeably, as well as in J5. The polarization is
$\sim 10\%$ in Knot J2. This conforms with the expectation for a shock, which partially
aligns a chaotic ambient magnetic field along the direction of the shock front.

\subsubsection{Moving Shock Model}

Moving shocks, caused by major disturbances in the injection at the central engine,
are a viable alternative to stationary ones, since the electron energies
and densities, as well as the bulk Lorentz factor, all increase behind the shock front
relative to those of the undisturbed flow. Such shocks are bounded by rarefactions on the side
toward the nucleus, since the shocked flow is faster than the undisturbed flow that was later
injected into the jet. We note that this scenario is in fact similar to the intermittent jet
proposal by \citet{Sta04}, since it involves a major enhancement in the jet flow
velocity and/or injected energy density that persists for thousands of years. Although
the direction of elongation of the knots seen in Figure \ref{fig1} is not always transverse to the flow,
we will simplify by approximating that the shock front is perpendicular to the velocity vector.
In the rest frame of the shock front, marked by a prime, the ratio of density ahead of
(subscript ``2'') to that behind (``1'') the front equals
$n_2/n_1 = (\Gamma'_{\rm 1}\beta'_{\rm 1})/(\Gamma'_{\rm 2}\beta'_{\rm 2})$, where $\beta$ is the
velocity in light units \citep[e.g.,][]{Sok04}.
We can obtain a consistent solution if, in the (unprimed) rest frame of the host galaxy of OJ287,
$\Gamma_1 = 5$ for the pre-shocked ambient jet and $\Gamma_{\rm s} = 11$ for the shock front.
Relativistic addition of the velocities
gives $\Gamma_2 = 7.9$ for the shocked plasma if $\beta'_{\rm 2} =1/3$, as appropriate
for an ultra-relativistic equation of state \citep{kon80}. Then $n_2/n_1=2.5$ and
$\Gamma_{\rm 1}/\Gamma_{\rm 2}=1.6$, which is also approximately $\delta_{\rm 1}/\delta_{\rm 2}$.
If the electrons are heated only by compression, the IC/CMB intensity,
$I_{\rm IC/CMB} \propto N_0\Gamma^4 \propto n_{\rm e}\gamma_{\rm min}^{2\alpha}\Gamma^4$,
behind the shock is still $\sim 20$ times the ambient value.

Either a standing or moving shock model can therefore account for the knotty structure,
although an oblique standing shock is favored, since it also explains the deceleration and
curvature of the jet. Within the model, the slight displacement
of the radio and X-ray intensity peaks (Fig.\ \ref{fig7}) implies that either the magnetic field
or the density of electrons with $\gamma \sim 10^4$ is higher near the shock front than
30-50 kpc downstream. In the moving shock model, the region of enhancement in $B$ and/or $N_0$ would
need to be 30-50 kpc closer to the nucleus than the location of the shock front.
This could be caused by turbulence that develops and amplifies the magnetic field
behind the shock \citep{Miz10} and possibly also accelerates particles. The displacement of
the various knots from the jet axis implies that the shocks cover only a fraction of the
jet cross-section.

\subsection{Comment on Spine-Sheath Models}

The strong, one-sided extended X-ray jet in OJ287 is best explained if the flow velocity
remains highly relativistic at least out to hundreds of kiloparsecs from the nucleus. We
detect bright knots and ambient X-ray emission across $\sim 2''$ (9 kpc) on both sides of the
centroid of the jet along much of its length. The length-to width ratio within $8''$ of the
nucleus is only $\sim 2$, which becomes $\sim 30$ when de-projected, corresponding to a full
opening angle of $\sim 2^\circ$ when deprojected. \citep[{\bf We note that this is similar to the
opening angle of the parsec-scale jet derived by}][.]{J05}Therefore, either the spine is not an
ultra-narrow pencil beam or its direction varies, as suggested by the placements of the knots.
Furthermore, the similar opening angle on parsec and kiloparsec scales supports our conclusion
that the extended jet remains relativistic, since FR~1 jets that decelerate tend to
broaden considerably on such scales \citep{L99}.

\section{Conclusions}

Our {\it Chandra} observations have revealed quite strong X-ray emission from the extended
jet of the BL Lac object OJ287, which has the unbeamed radio luminosity of an FR~1 source.
The X-ray emission persists
out to hundreds of kiloparsecs from the nucleus, while weak radio emission can be seen beyond
1 Mpc. This demonstrates that the main requirement for strong extended X-ray emission and
an extremely long jet is a highly relativistic outflow, independent of the class of radio source.

The observed SED of the extended jet of OJ287 eliminates synchrotron emission as the source of the
X-rays unless
the electron energy distribution differs from those in other blazars by concentrating essentially
all of the energy within the TeV range. The highly relativistic flow speed observed in the
parsec-scale jet and the relativistic beaming implied by the one-sidedness of the
kiloparsec-scale jet support an IC/CMB model for the extended X-ray emission. Since the
relatively low-energy electrons involved have extremely long radiative lifetimes, the
knotty structure of the jet requires shocks that compress the plasma and rarefactions that lower the density again. The  model that explains the widest range of observed properties
involves a standing shock at an angle of $7^\circ$
to the upstream flow, which can explain the bending beyond $8''$ as well as the knot to
inter-knot intensity ratio and decrease in bulk Lorentz factor. The knotty structure can
also arise from variations in the jet flow on time-scales of
thousands of years. This results in moving shock waves followed by rarefactions.
The knots lie at various positions
relative to the jet axis, which implies that, rather than confinement to
an extremely narrow spine, the structure subtends a more normal (for
relativistic jets) full opening angle of $\sim 2^\circ$, similar that that found on parsec scales.

We selected OJ287 because of its extended, although very weak, radio jet and its high apparent 
superluminal motion on parsec scales. Exploration with {\it Chandra} of the extended X-ray emission
around other BL Lac objects with very weak kiloparsec-scale radio jets detectable with the
EVLA could discover more structures with high X-ray to radio flux ratios. Observations of
such sources would further our knowledge of the physical development of relativistic
jets as they propagate through and beyond their host galaxies.

\acknowledgments

We thank C. C. Cheung for help with the analysis of
our HST image, and E. S. Perlman for advice on planning the HST observations.
This material is based upon work supported by the National Aeronautics and Space Administration under
Chandra Guest Investigator grant no. GO8-9097X administered by the Smithsonian Astrophysical
Observatory, Hubble Space Telescope General Observer grant HST-GO-11344.01-A administered by
the Space Telescope Science Institute, and Spitzer Space Telescope Research Support Agreement
1326216 administered by the Jet Propulsion Laboratory, and by the U.S.
National Science Foundation under grant no. AST-0907893. The
VLA is an instrument of NRAO. The National Radio Astronomy Observatory is a facility of the
National Science Foundation operated under cooperative agreement by Associated Universities, Inc.

\begin{deluxetable}{lccrrr}
\singlespace
\tablecolumns{6}
\tablecaption{\bf Observations}
\tabletypesize{\footnotesize}
\tablehead{
\colhead{Telescope}&\colhead{Instrument}&\colhead{Band}&\colhead{$UT_{start}$}&\colhead{$UT_{end}$}&\colhead{Exposure} \\
\colhead{(1)}&\colhead{(2)}&\colhead{(3)}&\colhead{(4)}&\colhead{(5)}&\colhead{(6)}
}
\startdata
Chandra&ACIS-S&0.2-10 keV&2007-12-27 17:02:36& 2007-12-28 07:15:20&50.4 ks \\
HST&WFPC2&600.1 nm&2007-12-18 21:43:39&2007-12-19 01:03:39&2760 s \\
Spitzer&IRAC&3.6 $\mu$m&2007-11-25 23:30:45&2007-11-26 00:26:41&2592 s \\
Spitzer&IRAC&5.8 $\mu$m&2007-11-25 23:30:45&2007-11-26 00:26:41&2592 s \\
VLA&A-array&1.425 GHz&1993-01-04 05:30:00&1993-01-04 14:30:00& 170 m \\
VLA&A-array&4.710 GHz&2003-08-31 14:00:00&2003-08-31 23:00:00& 200 m \\
VLA&B-array&14.940 GHz&2007-11-08 12:00:00&2007-11-08 15:00:00& 150 m \\
\enddata
\label{tab1}
\end{deluxetable}

\begin{deluxetable}{lccccrcc}
\singlespace
\tablecolumns{8}
\tablecaption{\bf X-Ray and Radio Measurements}
\tabletypesize{\footnotesize}
\tablehead{
\colhead{Knot}& \colhead{$\Phi$(X-ray)\tablenotemark{1}} & \colhead{$F_\nu$(1.4 GHz)\tablenotemark{2}} & 
\colhead{$F_\nu$(4.8 GHz)\tablenotemark{2}} & \colhead{$F_\nu$(15 GHz)\tablenotemark{2}} &
\colhead{Dimensions\tablenotemark{3}} & \colhead{$\alpha$(X-ray)\tablenotemark{4}}
& \colhead{$\alpha$(radio)\tablenotemark{4}}
}
\startdata
J1& --- & $2.0\pm 0.2$ & --- & --- & $< 0.5$ & --- \\
J2& $6.95\pm0.40$ & $2.2\pm 0.3$ & --- & --- & $1.5\times 1.2$ & $0.44\pm0.21$ & --- \\
J3& $3.40\pm0.28$ & $< 0.15$ & --- & --- & $1.3\times 1.3$ & $0.56\pm0.40$ & --- \\
J4& $6.16\pm0.38$ & $2.9\pm 0.2$ & --- & --- & $1.3\times 1.5$ & $0.72\pm0.21$ & --- \\
J5& $7.06\pm0.44$ & $1.6\pm 0.4$ & --- & --- & --- & $0.68\pm0.22$ & --- \\
J1-J5\tablenotemark{5}& $ 30.6\pm0.88$ & $ 10.9\pm0.5$ & $ 4.13\pm0.43$
& $1.43\pm1.03$ & 1.3 & 0.61$\pm$0.06 & 0.8$\pm$0.1 \\
Outer\tablenotemark{6}& --- & $ 4.6\pm0.9$ & ---
& --- & --- & --- & --- \\
\enddata
\tablenotetext{1}{Photon counts from 0.2 to 6 keV; units: $10^{-3}$ ph s$^{-1}$; no value
for J1 because it is confused with the PSF of the nucleus}
\tablenotetext{2}{Flux density in mJy; signal-to-noise ratio is insufficient to determine
for individual knots at 4.8 and 15 GHz}
\tablenotetext{3}{Full width across feature at half intensity parallel $\times$ perpendicular
to axis, in arcseconds}
\tablenotetext{4}{Spectral index, defined such that flux density $F_\nu \propto \nu^{-\alpha}$;
signal-to-noise ratio is insufficient to determine $\alpha$ for some knots, especially
at the radio frequencies}
\tablenotetext{5}{Includes emission between knots}
\tablenotetext{6}{Region beyond detected X-ray emission out to last radio feature seen in
Figure \ref{fig5}; only apparent at 1.4 GHz }
\label{tab2}
\end{deluxetable}

\clearpage

\end{document}